\documentclass[aps,twocolumn,pra,superscriptaddress,showpacs,tightenlines]{revtex4-1}
\usepackage{amsmath}
\usepackage{graphicx}
\usepackage{epsfig}
\usepackage{amsfonts}
\usepackage{color}

\begin{document}

\newcommand{\nn}{\nonumber}
\newcommand{\ms}[1]{\mbox{\scriptsize #1}}
\newcommand{\msi}[1]{\mbox{\scriptsize\textit{#1}}}
\newcommand{\dg}{^\dagger}
\newcommand{\smallfrac}[2]{\mbox{$\frac{#1}{#2}$}}
\newcommand{\ket}[1]{| {#1} \rangle}
\newcommand{\bra}[1]{\langle {#1} |}
\newcommand{\kket}[1]{| {#1} \rangle\rangle}
\newcommand{\bbra}[1]{\langle\langle {#1} |}
\newcommand{\pfpx}[2]{\frac{\partial #1}{\partial #2}}
\newcommand{\dfdx}[2]{\frac{d #1}{d #2}}
\newcommand{\half}{\smallfrac{1}{2}}
\newcommand{\s}{{\mathcal S}}
\newcommand{\jord}{\color{red}}
\newcommand{\kurt}{\color{blue}}

\title{Exact Trajectory Control for the non-Markovian Quantum Systems }

\author{S. L. Wu }
\affiliation{School of Physics and Materials Engineering,
Dalian Nationalities University, Dalian 116600 China}

\author{W. Ma }
\email{mawei@dlnu.edu.cn}
\affiliation{School of Physics and Materials Engineering,
Dalian Nationalities University, Dalian 116600 China}

\date{\today}

\begin{abstract}
We propose a systematic scheme to engineer quantum states of a quantum
system governed by a time-convolutionless  non-Markovian master equation.
According to the idea of the reverse engineering, the general algebraic equation
to determine the control parameters, such as coherent and incoherent control
fields, are presented. Without artificially engineering the time-dependent decay
rates and persisting the environment-induced Lamb shifts, the quantum state can still be
transferred into the target state with a finite period of time along an arbitrary
designed trajectory in Hilbert  space strictly.  As an application, we apply our scheme
to a driven two-level non-Markovian system, and realize the instantaneous steady
state tracking and the complete population inversion with control parameters
which are available in experimental-settings.
\end{abstract}

\maketitle

\section{Introduction}

Driving quantum systems, especially open quantum systems, to desired target states
with very high fidelity is a central goal in quantum sciences and technologies,
to realize efficient and scalable devices beyond the current state of proof-of-principle
demonstrations \cite{Koch2019,Osnaghi2001,Hacohen2018}. For controlling
open quantum systems with Markovian dynamics, many enlightening schemes have been
proposed to control open quantum systems, such as the adiabatic steady state scheme
\cite{Venuti2017}, the shortcut to equilibration scheme \cite{Dann2019},
the dissipative steady state preparing scheme \cite{Zippilli2021,Zheng2021},
and the mixed-state inverse engineering scheme \cite{Wu2021}. These schemes
transfer the quantum state into the target steady state with a satisfactory fidelity.

But to engineer a quantum state of the non-Markovian quantum system is
a different matter. Due to the memory effects of the environment,
the next state of a non-Markovian quantum system is determined by
each of its previous states \cite{Liu2019}.  The
decay rates are time-dependent, and may temporarily acquire negative
values \cite{Zhang2012}. This negative decay rate pushes the information
(coherence or/and energy)  to flow back into the open quantum system after
the information dissipates into the environment \cite{Li2018}. Therefore, to drive
the non-Markovian quantum systems  into a desired  target state along
an exact and designable trajectory  definitely is a non-trivial task.

In this paper, we focus on this issue, and propose a control scheme for the
non-Markovian quantum systems, which are governed by the time-convolutionless master
equation \cite{Gasbarri2018}. By parameterizing the trajectory of the quantum state
from the initial state to the target state, the control parameters can be determined
by reversely engineering time-dependent control Liouvillians. In this way, the quantum
state of the non-Markovian quantum system is transfered into the target
state along the parameterized trajectory strictly.  It should be emphasized that,
since the spectrum density of the environment is difficult to be engineered
in the experiments \cite{Jorgensen2020}, we do not select the decay
rates as a means of incoherent control. Although
the time-dependent decay rates draw the quantum state out of the trajectory, our scheme
can eliminate this effect and keep the quantum state on the designed
trajectory. We exam our scheme by applying it on the quantum state engineering tasks
of a driven two-level non-Markovian system.  The instantaneous steady state
tracking and the complete population inversion are realized. In this scenario, the two-level non-Markovian
systems are not only kinematically controllable, but also dynamically controllable,
which is impracticable for the Markovian case.

This paper is organized as follows. In Sec.\ref{method}, we present the exact
trajectory control scheme for the non-Markovian quantum systems governed
by the time-convolutionless master equation. Taking a non-Markovian two-level
system as an example, the instantaneous steady state tracking and the population
inversion are considered in Sec.\ref{example}. And we show that, attributing to the
information backflows, the population can be completely transferred into the
excited state of the two-level system with available control parameters in
 experiments. Finally, we give conclusions and discussions in Sec.\ref{conclusion}.

\section{The Method} \label{method}

In this work, we consider open quantum systems, where the coupling to a
reservoir leads to a non-Markovian dynamics for the system density matrix $\rho(t)$,
described by a time-convolutionless master equation in the Lindblad form
\begin{eqnarray}
\partial _t \rho(t)&=&\hat{\mathcal L}[\rho]\nonumber\\
&=&-i[{H}(t),{\rho}]+\sum_\alpha\gamma_\alpha(t) \mathcal D[L_\alpha](\rho),\label{meq0}
\end{eqnarray}
where $H(t)$ is the Hamiltonian containing the coherent controls on the system
and the Lamb shifts induced by the coupling to the reservoir, while $\mathcal D
[L_\alpha]$ is the Lindbladian with a Lindblad operator $L_\alpha$,
\begin{eqnarray}
\mathcal D[L_\alpha](\rho)&=& 2\,L_{\alpha}(t){\rho}L_{\alpha}^{\dagger}(t) -
\{L_{\alpha}^{\dagger}(t)L_{\alpha}(t), {\rho}\}.\label{lind}
\end{eqnarray}
Each Lindblad operator $L_\alpha(t)$ is associated with a dissipation channel occurring
at the time-dependent rate $\gamma_\alpha(t)$.  We consider the case where
$H(t)$, $\gamma_\alpha(t)$ and $L_\alpha(t)$ are time-dependent. This kind of master
equation can be applied for examples to photonic quantum systems
\cite{Vega2017} and  mesoscopic electron-phonon systems \cite{Pereverzev2006}.

Since the time-convolutionless master equation is linear in $\rho(t)$, it is
convenient to describe this master equation as a superoperator formalism in Hilbert-Schmidt
space \cite{Minganti2018}, wherein the density matrix is represented by a
$N^2$-dimensional vector
\begin{eqnarray}
\kket{\rho(t)}=(\rho_0(t),\rho_1(t),...,\rho_{N^2-1}(t))\dg,
\end{eqnarray}
where $\rho_i(t)$ is the $i$-th component of $\kket{\rho(t)}$ with a time-independent
basis $B_i$ of the  Hilbert-Schmidt space satisfying $\rho_i(t)=\text{Tr}
[\rho(t)B_i]$. On the other hand, the Liouvillian superoperator becomes a $N^2\times N^2$
time-dependent supermatrix $\mathcal L(t)$ whose elements are given by
$\mathcal L_{ij}(t)=\text{Tr}[B_i^\dagger(\hat{\mathcal L}[B_j])]$. Then the master
equation as shown in Eq.(\ref{meq0}) now reads
\begin{eqnarray}
\partial _t \kket{\rho(t)}=\mathcal L (t)\kket{\rho(t)},\label{meql}
\end{eqnarray}
with the Liouvillian supermatrix
\begin{eqnarray}
\mathcal L (t)&=&-i[H\otimes I-I\otimes H^\text{T}]\nonumber\\
&+&\sum_\alpha\gamma_\alpha\left(2L_{\alpha}\otimes L_{\alpha}^{*}-L_{\alpha}^{\dagger}
L_{\alpha}\otimes I-I\otimes L_{\alpha}^{\text{T}}L_{\alpha}^*\right),
\end{eqnarray}
where $A^\text{T}$ denotes the transposition of the operator $A$ and $I$, is the
identity operator.

The aim of the control scheme is to transfer the quantum system from a known and
arbitrary initial state $\rho(0)$ to a desired target state $\rho(t_f)$ along a preset
trajectory. The choice on the bases of the  Hilbert-Schmidt is not unique, and the
principle on this choice is determined by how to simplify complexity for obtaining the
feasible control parameters in the Liouvillian superoperator. Without the loss of generality,
the basis set of the  Hilbert-Schmidt space can be chosen as the  SU($N$) Hermitian
generators $\{T_i\}_{i=1}^{N^2-1}$ and the identify operator $T_0\equiv I$.
Thus the density matrix can be expanded by these bases, and yields
\begin{eqnarray}
| \varrho (t)\rangle\rangle=\frac{1}{N}\left ( | I\rangle\rangle+\sqrt{
\frac{N (N-1)}{2}}\sum_{i=1}^{N^2-1} r_i |T_i\rangle\rangle \right),\label{rho}
\end{eqnarray}
where $\vec r= (r_1, r_2, ..., r_{N^2-1}) $ is the generalized Bloch vector with $\sum_i |r_i|^2<1$.
Within this notation, the density matrix can be parameterized by
$N^2-1$ independent coefficients.

On the other hand,  the Liouvillian superoperator contains all of the
control parameters which can be applied in the real-world
experimental setting. The control on the quantum system comes from two
manners, i.e., the coherent control and the incoherent control.
The coherent controls on the quantum system applied in the experiment
are contained in the Hamiltonian of the Liouvillian superoperator. By using the SU($N$)
Hermitian generators $\{T_i\}_{i=0}^{N^2-1}$ ( $T_0\equiv I$ is
the identify operator), the Hamiltonian can be expressed as
\begin{eqnarray}
H (t) =\sum_{i=0}^{N^2-1}c_i(t)\,T_i,\label{hami}
\end{eqnarray}
where $c_i(t)$ denotes the control parameter for the coherent operation $T_i$ on
the system. The incoherent controls come from the couplings to the environment,
which are reflected in the master equation by the Linbladian. Generally, the Lindblad
operators can be written as superpositions of the SU($N$) Hermitian generators,
such as
\begin{eqnarray}
L_{\alpha}(t)=\sum_{j=1}^{N^{2}-1}l_{j}^{(\alpha)}(t)\,T_{j}
\end{eqnarray}
with complex expansion coefficients $l_{j}^{(\alpha)}(t)$. Here we assume that
these complex coefficients $\{l^{(\alpha)}_j(t)\}$ include incoherent control
parameters which  are tunable in experiment and influence the system via incoherent
manners.These incoherent control parameters include, but not limited to,
the main excitation numbers of the environment \cite{Basilewitsch2019,Shabani2016}, the correlation of the
environment \cite{Seetharam2021}, even  extra noises \cite{Pechen2016}. As the
restriction on our scheme, the correlation functions of the environment are invariant.
Thus the decay rates and the  Lamb shifts caused by the interaction between
the open quantum systems and the environments can not be changed artificially,
which distinguishes our scheme from pervious schemes on this topic \cite{Alipour2020}.

Here we are in the position to determine all of the control parameters (coherent
and incoherent) reversely. In fact, the density operator vector is the solution of
Eq.(\ref{meql}). Our scheme is to preset the density operator $\rho(t)$, and then
to determine the control parameters $\{c_i(t),l^{(\alpha)}_j(t)\}$ by Eq.(\ref{meql}).
At the beginning, we parameterize the density operator by the generalized Bloch
vector as shown in Eq.(\ref{rho}). The initial and final Bloch vectors have to correspond
to the initial and target state of the control task. Thus the time-dependent Bloch
vector corresponds to a trajectory of the quantum state in the Hilbert space, which
connects the initial state and target state. Then we deal with the Liouvillian
supermatix. The elements of the Liouvillian supermatrix can be determined by
$\mathcal{L}_{ij}(t)=\text{Tr}[T_{i}^{\dagger}(\hat{\mathcal{L}}[T_{j}])]$. In order
to distinguish the coherent and incoherent control manners, we divide the
Liouvillian supermatrix into three parts. Thus, we rewrite the
time-convolutionless master equation in components of the generalized Bloch vector,
\begin{eqnarray}
\partial_{t}r_{i}(t)	=	\sum_{j=1}^{N^{2}-1}\left(\mathcal{C}_{ij}+
\mathcal{I}_{ij}\right)\,r_{j}(t)+\mathcal L^0_i.\label{beq}
\end{eqnarray}
The coherent part reads
\begin{eqnarray}
\mathcal{C}_{ij}=\sum_{k=1}^{N^{2}-1}c_{k}(t)\,\frac{f_{kji}}{2}
\end{eqnarray}
and the incoherent part takes the form
\begin{eqnarray}
\mathcal{I}_{ij}=\sum_{m,n=0}^{N^{2}-1}\left(\sum_{\alpha}\gamma_{\alpha}
l_{m}^{(\alpha)}(t)l_{n}^{(\alpha)*}(t)\right)s_{mn,ji}
\end{eqnarray}
with
\begin{eqnarray}
s_{mn,ji}&=&\frac{1}{2N}\left(\delta_{im}\,\delta_{jn}-\delta_{mn}\,
\delta_{ij}\right)\nonumber\\
&&+\frac{1}{4}\sum_{k=1}^{N^{2}-1}\left(\left(i\,f_{jnk}+d_{jnk}\right)
\left(i\,f_{imk}+d_{imk}\right)\right.\nonumber\\
&&\left.-\left(i\,f_{mnk}+d_{mnk}\right)\,d_{kji}\right),
\end{eqnarray}
where $f_{ijk}$ and $d_{ijk}$ are the structure constants and the d-coefficients
of the SU($N$) Lie algebra, respectively. Moreover, the last terms in Eq.(\ref{beq})
can be written as
\begin{eqnarray*}
\mathcal{L}_{k}^{0}
&=&\sum_\alpha \gamma_\alpha(t)\left(\sum_{i,j=1}^{N^{2}-1}l_{i}^{(\alpha)}(t)
l_{j}^{(\alpha)*}(t)g_{ijk}\right),
\end{eqnarray*}
with $g_{ijk}=\left(\left(i\,f_{jik}+d_{jik}\right)-\left(i\,f_{ijk}+d_{ijk}\right)\right)$.
The derivation of the coherent and incoherent parts of the Liouvillian supermatrix can be
found in Appendix \ref{appendix}.

In fact, Eq.(\ref{beq}) is not only linear to the components of the Bloch vector,
but also linear to the control parameters $\{c_i(t),\sum_\alpha \gamma_\alpha
l^{(\alpha)}_i(t)l^{(\alpha)*}_j(t)\}$. Here, we further assume
that there is only one tunable incoherent control parameter in every Lindbladian
$\mathcal D[L_\alpha]$, i.e., $l^{(\alpha)}_i(t)=\sqrt{\tilde c_\alpha(t)}\tilde l^{(\alpha)}_i$,
where  $\tilde c_\alpha (t)$ is a real incoherent control parameter and
$\{\tilde l^{(\alpha)}_i\}$ are  time-independent expansion coefficients.
Thus, we have $\sum_\alpha \gamma_\alpha l^{(\alpha)}_i(t)l^{(\alpha)*}_j(t)=\sum_\alpha
\gamma_\alpha \tilde c_\alpha(t)\tilde l^{(\alpha)}_i\tilde l^{(\alpha)*}_j$.
In this notation, the equations of the control parameters are given by
\begin{eqnarray}
\sum_j \Lambda^c_{ij}c_j+\sum_{\alpha}\gamma_{\alpha}\left(\Lambda^i_{\alpha\,i}
+\Lambda^0_{\alpha\,i}\right)\tilde c_{\alpha}-\partial_t r_i(t)=0,\,\forall i\nonumber\\\label{leq}
\end{eqnarray}
where the coefficient matrixes for the coherent control and incoherent parameters are
\begin{eqnarray}
\Lambda^c_{ij}&=&\sum_{k=1}^{N^{2}-1}r_{k}(t)\,\frac{f_{jki}}{2},\nonumber\\
\Lambda^i_{\alpha\,i}&=&\sum_{j,m,n=0}^{N^{2}-1} r_j(t)\left(\tilde l_{m}^{(\alpha)}(t)
\tilde l_{n}^{(\alpha)*}(t)\right)s_{mn,ji},\nonumber\\
\Lambda^0_{\alpha\,i}&=&\sum_{j,k=1}^{N^{2}-1}\tilde l_{j}^{(\alpha)}(t)
\tilde l_{k}^{(\alpha)*}(t)g_{jki}
\end{eqnarray}
In order to obtain the control parameters $\{c_i(t),\tilde c_{(\alpha)}(t)\}$,
we need to solve above equations.

As we see, Eq.(\ref{leq}) can not provide the single unique solution of the control
parameters in general.  In practice, not all of control can be applied on
the system.  For instance, a $\Delta$-type coherence control can be realized in
artificial structure but not in real atoms via dipole-dipole coupling due to the
selection rule \cite{Chen2010}. Also, the deocoherence channels used in the scheme
must be restricted by the real-world setting. In other words, the open
quantum systems must be dynamically controllable \cite{Wu2007,Grigoriu2013}.
Therefore, the selection on control parameters in our scheme, two principles have
to be stuck up: (i). The number of the control parameters is equal to equation number in Eq.(\ref{leq}),
which ensures the single unique control parameter in the control scheme; (ii).
All of the control technologies corresponding to the control parameters
have to be available in the real-experiment setting. To meet above requirements,
the control technologies with corresponding control parameters have to be
selected not only on experimental conditions in the laboratory, but also the
symmetry of the open quantum systems \cite{Lostaglio2017,buvca2012}.

\section{Applications: A Two-level non-Markovian system}\label{example}

We consider a two-level system with the transition frequency $\omega_0$ driven
by an external laser of frequency $\omega_L$ \cite{Shen2014,Cui2008}. There
is a detuning $\Delta= \omega_0-\omega_L$ between the two-level system and the
external laser. The two-level atom is embedded in a bosonic reservoir at a
finite temperature $T$. In a rotating frame, the Hamiltonian can be written as
 \begin{eqnarray}
H=H_s+H_e+H_i,
 \end{eqnarray}
with
 \begin{eqnarray}
H_s&=&\Delta\sigma_+\sigma_-+\Omega(t)\sigma_++\Omega^*(t)\sigma_-,\nonumber\\
H_e&=&\sum_k  \Omega_k a_k^\dagger a_k,\nonumber\\
H_i&=&\sum_k g_k\sigma_+ a_k +\text{h.c}.,
 \end{eqnarray}
where $\Delta=\omega_0-\omega_L$ $\Omega_k=\omega_k-\omega_L$, $\sigma_+=
\ket{e} \bra{g}$, $\Omega (t)=\Omega_x(t)+i\Omega_y(t)$ is the
time-dependent control field, h.c. stands for the Hermitian conjugation,
$a_k$ and $g_k$ stand for the annihilation operator and coupling constant, respectively.

By the atomic coherent-state path-integral method\cite{Shen2014}, an exact non-Markovian
master equation can be obtained to describe  the dynamics of the open two-level system,
 \begin{eqnarray}
\partial_t \rho(t)&=&\hat{\mathcal L}_0(t)\rho(t)\nonumber\\&=&-i
[H_s^{\text{R}}(t),\rho(t)]+\Gamma_0 (N+1)\hat{\mathcal D}[\sigma_-](\rho(t))\nonumber\\
&&+\Gamma_0 N\hat{ \mathcal D}[\sigma_+](\rho(t)),\label{nmeq}
 \end{eqnarray}
with the effective Hamiltonian,
 \begin{eqnarray}
H_s^{\text{R}}(t)=s_0(t)\sigma_+\sigma_-+\Omega^\text{R}(t)\sigma_++
\Omega^{\text{R}*}(t)\sigma_-.
 \end{eqnarray}
$s_0(t)$ and $\Omega^\text{R}(t)$ are the Lamb shift and the renormalized
driving field respectively, which are resulted by memory effects of the bosonic
reservoir. The time-dependent decay rate $\Gamma_0(t)$ describes the
dissipative non-Markovian dynamics due to the interaction between the system
and environment. $N=[\exp(\hbar\omega_0/k T_0)-1]^{-1}$ stands for the mean
excitation number. They are both associated with the spectral density  and the
temperature $T_0$ of the reservoir. All of these time-dependent coefficients
can be given explicitly as follows
\begin{eqnarray}
s_0(t)&=&-\text{Im}\left[\frac{\partial_t u(t)}{u(t)}\right],\!
\Gamma_0(t)=-\text{Re}\left[\frac{\partial_t u(t)}{u(t)}\right],\label{lamdecay}\\
\Omega^R(t)&=&i\left[\partial_t h(t)-h(t)\frac{\partial_t u(t)}{u(t)}\right],
 \end{eqnarray}
 where Re[$\bullet$] and Im[$\bullet$] represent the real and imaginary
 part of the argument, respectively. $u(t)$ and $h(t)$ satisfy the following
 equations
\begin{eqnarray}
\partial_tu(t)+i\Delta u(t)+\int_0^t f(t-t') u(t')dt'&=&0,\label{e}\\
\partial_th(t)+i\Delta h(t)+\int_0^t f(t-t') h(t')dt'&=&-i\Omega,\label{eh}
 \end{eqnarray}
with
\begin{eqnarray}
f(t-t')=\int d\omega\,J(\omega)\exp(-i(\omega-\omega_L)(t-t')),\label{kernel}
 \end{eqnarray}
and the boundary conditions $u(0)=1$, $h(0)=0$. We assume that the spectral
density of the bosonic reservoir has a Lorentzian form \cite{Shen2013,Shen2014,Haikka2010}
 \begin{eqnarray}
J(\omega)=\frac{\gamma_0}{2\pi}\frac{\lambda^2}{(\omega-\omega_0+\delta)^2+\lambda^2},\label{specden}
 \end{eqnarray}
where $\delta=\omega_0-\omega_c$ is the detuning of $\omega_c$ to $\omega_0$,
$\omega_c$ is the center frequency of the cavity, and $\lambda$ is the
spectral width of the reservoir. The parameter $\gamma_0$  is the decoherence
strength of the system in the Markovian limit with a flat spectrum. Substituting
Eq.(\ref{specden}) into Eq.(\ref{kernel}), we obtain the two-time correlation functions
 \begin{eqnarray}
f(t-t')=\frac{1}{2}\lambda\Gamma\exp[-(\lambda+i\Delta-i\delta)(t-t')].
 \end{eqnarray}
Thus, the solutions of Eq.(\ref{e}) and Eq.(\ref{eh})  take the forms
 \begin{eqnarray}
u(t)&=&k(t)\left[\cosh\left(\frac{dt}{2}\right)+\frac{\lambda+i\delta}{d}
\sinh\left(\frac{dt}{2}\right)\right],\\
h(t)&=&-i\,\int_0^t \Omega(t')u(t-t')\,dt',
 \end{eqnarray}
 where $k(t)=\exp(-(\lambda+2i\Delta-i\delta)t/2)$ and $d=
 \sqrt{(\lambda-i\delta)^2-2\gamma_0\lambda}$.

To reversely engineer the non-Markovian two-level system, we parameterize
the quantum state by a Bloch vector, which can be written as
  \begin{eqnarray}
| \mathcal \varrho(t)\rangle\rangle=\frac{1}{2}\left(| I \rangle\rangle+
\sum_{i=x,y,z} r_i |\sigma_i \rangle\rangle \right),
 \end{eqnarray}
where $r_i$ is the $i$-th component of the Bloch vectors, and $\sigma_i$
is $i$-component of the Pauli operators. Thus, the quantum state of the two-level
system has three independent parameters. And the effective Hamiltonian can be rewritten as
\begin{eqnarray}
H_s^\text{R}(t)=s_0(t)\sigma_+\sigma_-+\Omega^\text{R}_x(t)\sigma_x+
\Omega^{\text{R}}_y(t)\sigma_y.
 \end{eqnarray}
We have assumed that the spectrum density is untunable in
experimental-settings, so the decay rate $\Gamma_0(t)$ can not
be a candidate for the incoherent control parameters. Therefore,
the coherent control parameters are chosen as $\Omega^
\text{R}_x(t)$ and $\Omega^\text{R}_y(t)$, while the main excited number $N(t)$
acts as the incoherent control parameter. Taking the control parameters and
the components of the Bloch vector into Eq.(\ref{beq}), it yields
  \begin{eqnarray}
\mathrm{\dot{r}_{x}}&=&2\,\Omega_{y}^{\text{R}}\,\mathrm{r_{z}}-s_{0}\,
\mathrm{r_{y}}-\left(2\,N+1\right)\,\Gamma_{0}\,\mathrm{r_{x}},\nonumber\\
\mathrm{\dot{r}_{y}}&=&s_{0}\,\mathrm{r_{x}}-2\,\mathrm{\Omega_{x}^{\text{R}}}\,
\mathrm{r_{z}}-\left(2\,N+1\right)\,\Gamma_{0}\,\mathrm{r_{y}},\nonumber\\
\mathrm{\dot{r}_{z}}&=&2\,\mathrm{\Omega_{x}^{\text{R}}}\,\mathrm{r_{y}}-2\,
\mathrm{\Omega_{y}^{\text{R}}}\,\mathrm{r_{x}}-2\,\Gamma_{0}\,\left(\left(2\,N+
1\right)\,\mathrm{r_{z}}+1\right)\!\!,\label{sr}
 \end{eqnarray}
where $\dot r_i$ denotes the time-derivative of the $i$-th component
of the Bloch vector. For the sake of brevity, we also ignored "(t)". The goal
of the reverse engineering scheme is to find
the control parameters which drive the two-level system to evolve as
users prescribe.  To achieve this goal, we reversely solve Eq.(\ref{sr}), and obtain
 \begin{eqnarray}
\Omega_{x}^{\text{R}}	&=&	\frac{\left(\mathrm{r}^{2}+\mathrm{r_{z}}^{2}\right)\,\left(\mathrm{r_{x}}
\,s_{0}-\mathrm{\dot{r}_{y}}\right)+\left(\vec{\mathrm{r}}\cdot
\dot{\vec{\mathrm{r}}}+2\,\Gamma_{0}\,\mathrm{r_{z}}\right)\,\mathrm{r_{y}}}
{2\,\mathrm{r_{z}}\,\left(\mathrm{r}^{2}+\mathrm{r_{z}}^{2}\right)},\nonumber\\
\Omega_{y}^{\text{R}}&	=&	\frac{\left(\mathrm{r}^{2}+\mathrm{r_{z}}^{2}\right)\,\left(\mathrm{r_{y}}\,
s_{0}+\mathrm{\dot{r}_{x}}\right)-\left(\vec{\mathrm{r}}\cdot
\dot{\vec{\mathrm{r}}}+2\,\Gamma_{0}\,\mathrm{r_{z}}\right)\,\mathrm{r_{x}}}
{2\,\mathrm{r_{z}}\,\left(\mathrm{r_{x}}^{2}+\mathrm{r_{y}}^{2}+2\,\mathrm{r_{z}}^{2}
\right)},\nonumber\\
N	&=&	-\frac{2\,\Gamma_{0}\,\mathrm{r_{z}}+\vec{\mathrm{r}}\cdot\dot{\vec{\mathrm{r}}}
+\mathrm{\Gamma_{0}}\,\left(\mathrm{r}^{2}+\mathrm{r_{z}}^{2}\right)}
{2\,\mathrm{\Gamma_{0}}\,\left(\mathrm{r}^{2}+\mathrm{r_{z}}^{2}\right)},\label{peq}
 \end{eqnarray}
with $\mathrm{r^2=r_x^2+r_y^2+r_z^2}$ and $\vec{ \mathrm{r}}\cdot
\dot{\vec{ \mathrm{r}}}=\mathrm{r_x\dot r_x+r_y\dot r_y+r_z\dot r_z}$.

In case of the Markovian dynamics, the Lamb shift vanishes
($s_0=0$) and the decay rate is time-independent ($\Gamma_0=\gamma_0$).
$\Omega_{x,y}^{\text{R}}$ are the control fields acting on the two-level
system. Therefore, the set of control parameters proposed in Eq.(\ref{peq}) is
a control protocol for two-level systems in a Markovian environment. In other words, our scheme is also an  available option for controlling
Markovian quantum systems. We can rewrite the last equation in Eq.(\ref{peq}) as $2\,\Gamma_{0}\,\mathrm{r_{z}}+\vec{\mathrm{r}}\cdot\dot{\vec{\mathrm{r}}}=
-\mathrm{(2N+1)}\,\left(\mathrm{r}^{2}+\mathrm{r_{z}}^{2}\right)\Gamma_{0}$.
By substituting it into the expression of $\Omega_{x,y}^{\text{R}}$,
we obtain the same control field as that used in Ref.\cite{ran2020}. Moreover,
we may set that the main excitation number $N$ is invariant in the control process,
which is the very  constraint condition mentioned in  Ref.\cite{ran2020}.

Here we want to emphasize that it is not the only choice for the control
parameters as used above. For instance, while we keep the incoherent control protocol
invariant, the detuning $\Delta$ can also be selected as a coherent control parameter,
which will provide another control protocol without using the control parameter
$\Omega^{\text{R}}_y$ (see Appendix \ref{appendixb}). In fact, whether
coherent or incoherent, as long as the solutions of Eq.(\ref{leq}) exist,
these control parameters can be candidates for control protocols. It means
that the two-level system is kinematically controllable for our control
scheme \cite{Klaus2006,Wu2007}. If the control protocol is totally coherent, it can be
verified that Eq.(\ref{leq}) has no solution, which indicates that the
open two-level systems are kinematically incompletely controllable for the pure coherent control
protocol \cite{Koch2016,Wu2007}. On the other hand, there are always
restrictions on controls, such as the finite pulse strength and detuning,
nonnegative main excitation  numbers. Thus, although the system is
kinematically controllable with the proper control protocol, it still can
not be realized in real-experimental setting. In other words,  an open quantum
system which is kinematically  controllable is not always dynamically controllable
by using the available  set of controls \cite{Koch2016}.
As shown in Eq.(\ref{peq}), the control parameters relate to the trajectory
of the quantum state in the Hilbert space (components of
the Bloch vector). Therefore, our scheme  can enhance the dynamical
controllability by designing proper trajectories of the quantum states
in the Hilbert space.

\subsection{The Steady State Tracking}

In this subsection, we drive the two-level non-Markovian system to track
the instantaneous steady state of a particular reference Liouvillian $\mathcal L_0(t)$
\cite{Wu2019}, which  is often used in the quantum thermodynamics \cite{cavina2017,anders2017}
and the quantum many-body theory \cite{Venuti2017}.
In particular, to transfer the quantum state of open quantum systems along the
instantaneous steady state strictly is critical for optimizing the performance
of the quantum heat engine \cite{miller2021,raja2021}.

Let the reference Liouvillian $\hat{\mathcal L}_0(t)$ take the same form as the
non-Markovian master equation presented in Eq.(\ref{nmeq}) with  a reference
Hamiltonian $$H_0(t)=s_0(t)\sigma_+\sigma_-+\Omega^\text{R}_0(t)\sigma_x,$$ and
a constant main excitation number $N_0$. Thus, the reference Liouvillian
supermatrix reads
 \begin{eqnarray}
&&{\mathcal L}_0(t)=\Gamma_0\times\nonumber\\
&&\left(\begin{array}{cccc}
-(N' + 1)& i\frac{{\Omega^\text{R}_{0}}}{\Gamma_{0} } & -i\frac{{\Omega^\text{R}_{0}}}{\Gamma_{0} }& N'-1\\
i\frac{{\Omega^\text{R}_{0}}}{\Gamma_{0}}& -N'-i\frac{{s_{0}}}{\Gamma_{0}}& 0 & -i\frac{{\Omega^\text{R}_{0}}}{\Gamma_{0}}\\
-i\frac{{\Omega^\text{R}_{0}}}{\Gamma_{0} } & 0 & -N'+i\frac{{s_{0}}}{\Gamma_{0} } & i\frac{{\Omega^\text{R}_{0}}}{\Gamma_{0}}\\
(N’ + 1)& -i\frac{{\Omega^\text{R}_{0}}}{\Gamma_{0} } & i\frac{{\Omega^\text{R}_{0}}}{\Gamma_{0} } & -N'+1
\end{array}\right)\nonumber
 \end{eqnarray}
with $N'=2N_0+1$. The instantaneous steady state of two-level system is given
by the condition ${\mathcal L}_0(t) |\rho_0(t)\rangle\rangle=0$, which yields
  \begin{eqnarray}
|\rho_0\rangle\rangle=\frac{1}{z}
\left(\begin{array}{c}
N_0\,\left(N'^{2}\Gamma_{0}^{2}+s_{0}^{2}\right)+N'{\Omega^\text{R}_{0}}^{2}\\
\left(iN'\Gamma_{0}-s_{0}\right)\Omega^\text{R}_{0}\\
-\left(iN'\Gamma_{0}+s_{0}\right)\Omega^\text{R}_{0}\\
\left(N_0+1\right)\left(N'^{2}\Gamma_{0}^{2}+s_{0}^{2}\right)+N'{\Omega^\text{R}_{0}}^{2}
 \end{array}\right)\label{ssn}
\end{eqnarray}
with  the factor $z=N'\left(\Gamma_{0}^{2}N'^{2}+s_{0}^{2}+2{\Omega^\text{R}_{x}}^{2}\right) $.

We impose that the initial and final Bloch vectors are the very Bloch vectors for the
instantaneous steady state $|\rho_0(t)\rangle\rangle$ (Eq.(\ref{ssn})) at $t=0$ and
$t=t_f$. Since there is not adiabatic theorem for the non-Markovian case,  the reference
Liouvillian $\hat{\mathcal L}_0(t)$  can not drive the quantum system into the
final steady state along the instantaneous steady state, even if $\dot{\Omega}^{\text{R}}_0
\rightarrow 0$. Hence, it is not necessary to compel $\Omega_x^{\text{R}}(t)
=\Omega^{\text{R}}_0(t)$ at the initial and final moment. What we need to concern is to
find a set of proper control parameters which ensures that the quantum state
tracks the instantaneous steady state trajectory strictly. The instantaneous steady state
Eq.(\ref{ssn}) can be rewritten in the form of the Bloch vector as
 \begin{eqnarray}
 &&r_x(t)=-\frac{2}{z}\,\Omega^{\text{R}}_0(t)s_0(t),\nonumber\\
&&r_y(t)=-\frac{2}{z}\,N'\Omega^{\text{R}}_0(t)\Gamma_0(t),\nonumber\\
&&r_z(t)=-\frac{1}{z}\,\left(s_0(t)^2+N'^2\Gamma_0(t)^2\right).\label{bvs}
 \end{eqnarray}
We suppose that the  reference control field $\Omega_0^{\text{R}}(t)$ tunes up from 0 to
a finite strength $\Omega_c$, and the time  derivative of $\Omega
^{\text{R}}_0(t)$  as zero at the initial and final instant. Therefore, we assume the
following time-dependent profile of $\Omega^{\text{R}}_0(t)$
\begin{eqnarray}
{\Omega_0^{\text{R}}(t)=6\Omega_c \,\frac{t^2}{t_f^2}\left(\frac{1}{2}-\frac{t}{3t_f}\right).}\label{ome0}
\end{eqnarray}
Substituting Eqs.(\ref{lamdecay}) and (\ref{ome0}) into Eq.(\ref{peq}), we can obtain all
analytical expressions  for the control parameters, which can drive quantum
state into the target steady state along the instantaneous steady state strictly.

 \begin{figure}[htbp]
\centerline{\includegraphics[width=1.1\columnwidth]{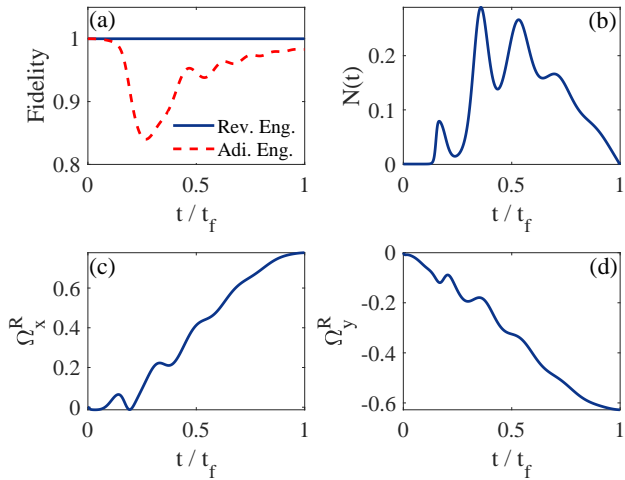}}
\caption{(a) The  fidelity for the reverse  engineering protocol (the blue solid line) and the adiabatic engineering protocol (the red dash line)  vs the
dimensionless time $t/t_f$. The control parameters ((b) the main excitation number $N$, (c) the coherent control field $\Omega^{\text{R}}_x$, and (d) the coherent control
field $\Omega^{\text{R}}_y$) as a function of the  dimensionless time $t/t_f$. Parameters: $\lambda=0.5\gamma_0,\,\Delta=0.1\gamma_0,\,\delta=0.5\gamma_0,\,
\Omega_c=10\gamma_0, t_f=10/\gamma_0$, and $N_0=10^{-5}$.  \textbf{We set $\gamma_0=1$ as the unit of $\Omega^{\text{R}}_x$ and
$\Omega^{\text{R}}_y$.}}\label{nm2l}
\end{figure}

As shown in FIG.\ref{nm2l} (a), the evolutions of the fidelities between
the quantum state governed by the master equation Eq.(\ref{nmeq}) and the
instantaneous steady state given by Eq.(\ref{ssn}) are plotted for the inverse
engineering protocol (blue solid line) and the adiabatic  engineering protocol
(red dash line). For the reverse engineering scheme, the quantum state of the
open two-level system strictly follows the instantaneous steady state.
When the adiabatic engineering protocol is used, i.e., $\Omega^{\text{R}}_x=
\Omega^{\text{R}}_0$ and $\Omega^{\text{R}}_y=0$, the fidelity decrease evidently. Even
the performance of the adiabatic engineering protocol is satisfied  as long as the
control time length $t_f$ increases,the quantum state deviates from the steady
state trajectory at the intermediate time due to the rapid oscillation of the
decay rate $\Gamma_0(t)$ and the Lamb shift $s_0(t)$.

The main excitation number $N(t)$ and the control field $\Omega^{\text{R}}_{x,y}(t)$
are also plotted in FIGs.\ref{nm2l}(b), (c) and (d). On the one hand, all of the control
parametersoscillate with time, which is essential to offset effect of  the
rapid oscillation of the decay rate $\Gamma_0(t)$ and the Lamb shift $s_0(t)$. In this way,
the quantum state is suppressed on the instantaneous steady state. On the other hand, due to
the nonzero Lamb shift $s_0(t)$, the coherent control field $\Omega^{\text{R}}_{y}(t)$ is needed
in the reverse engineering protocol, which does not appear in the reference Hamiltonian
(or Liouvillian). If $s_0(t)=0$, the $x$-th component of the Bloch vector will be zero,
which will result in the absence of coherent control field $\Omega^{\text{R}}_{y}(t)$
(see Eqs. (\ref{peq}) and (\ref{bvs})). This is the significant difference from the Markovian
reverse engineering protocol counterpart.

\subsection{The Population Inversion}

Similar ideas can be applied to the population inversion of the two-level open quantum state.
For convenience, we express the Bloch vector $\vec r$ by means of spherical polar coordinates,
i.e.,
 \begin{eqnarray}
 &&r_x=r\,\sin\theta\,\sin\phi,\nonumber\\
&&r_y=r\,\cos\theta\,\sin\phi,\nonumber\\
&&r_z=r\,\cos\phi.\label{refs}
 \end{eqnarray}
Our aim is to transfer the quantum state from the ground state $\ket{0}$ into the excited
state $\ket{1}$.  $\ket{0}$ and $\ket{1}$ are the eigenvectors of $\sigma_z$. For that,
we set the boundary conditions of quantum state as $\phi(0)=\pi$, $r(0)=1$, $\phi(t_f)=0$,
and $r(t_f)=1$. It is free to choose the values of $\theta(0)$ and $\theta(t_f)$.
According to Eq.(\ref{peq}), when $\phi\rightarrow \pi/2$,  the coherent control fields
$\Omega^{\text{R}}_{x}$ and $\Omega^{\text{R}}_{y}$ tend to be infinite. In order to eliminate this singularity,
we require $s_0=0$, $\dot r =0$ and $\dot\theta=0$ for $\phi= \pi/2$. Here we should
mention that the requirement $s_0(t_i)=0$ for some intermediate moment
$t_i$ can be realized by picking a proper detuning $\Delta$. In addition, for
the non-Markovian dynamics of open quantum systems, the decay rates are negative in
some intermediate duration. Thus the main excitation $N(t)$ will be infinite at the
moment for $\Gamma_0=0$. Yet, if we require $\vec r\cdot \dot{\vec r}=0$ at this
point, a reasonable main excitation number can be obtained (see Eq.(\ref{peq})).

 \begin{figure}[htbp]
\centerline{\includegraphics[width=1.1\columnwidth]{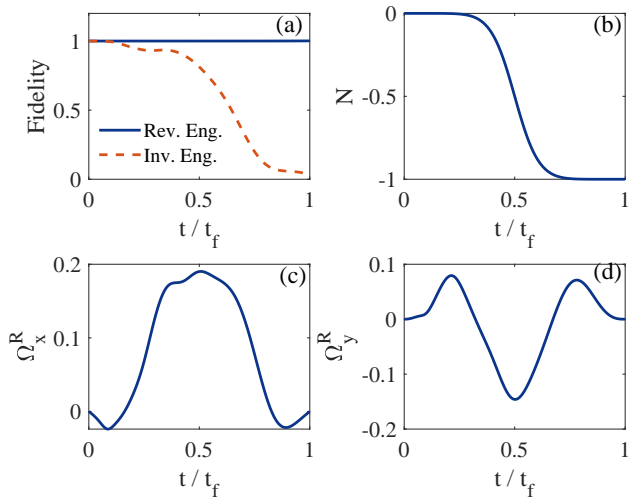}}
\caption{(a) The evolution of the fidelity of the reverse  engineering protocol
(the blue solid line) and the adiabatic engineering protocol (the red dash line).
The control parameters ((b) the main excitation number $N$, (c) the coherent
control field $\Omega^{\text{R}}_x$, and (d) the coherent control
field $\Omega^{\text{R}}_x$) as a function of the  dimensionless time $t/t_f$.
Parameters: $\lambda=0.5\gamma_0,\,\Delta=0.1\gamma_0,\,\delta=0.5\gamma_0,\,
\Omega_c=10\gamma_0, t_f=10/\gamma_0$ and $N_0=10^{-5}$.  We set
$\gamma_0=1$ as the unit of $\Omega^{\text{R}}_x$ and
$\Omega^{\text{R}}_y$.}\label{pures}
\end{figure}

Firstly, we show that the population inversion with a pure-state trajectory is kinematically
controllable, but not dynamically controllable.  To interpolate at intermediate times,
we consider a polynomial ansatz of $\theta$ and $\phi$ as a function of time $t$,
\begin{eqnarray}
r(t)&=&1,\nonumber\\
\phi(t)&=&\pi\frac{t^2}{t_f^2}\left(3-2\frac{t}{t_f}\right),\nonumber\\
\theta(t)&=&\theta\left(\frac{t_f}{2}\right)\frac{t^2}{t_f^2}\left(1-\frac{t}{t_f}\right)^2,
\end{eqnarray}
with $\theta(0)=\theta(f_f)=0$ and an arbitrary $\theta\left(\frac{t_f}{2}\right)$ at $t=t_f/2$.
Under this ansatz, for $t=t_f/2$, we have $\phi(t_f/2)=\pi/2$ and $\dot \theta (t_f/2)=0$,
which result in a reasonable coherent control fields in the control period. Figure \ref{pures} (a)
shows the fidelities between the quantum state $\rho(t)$ and the preset trajectory given by
Eq.(\ref{refs}) for the reverse engineering protocol (the blue solid line) and the inverse
engineering protocol of closed quantum systems. The control parameters are plotted
in FIG. \ref{pures} (b), (c) and (d). As we see, the reverse engineering protocol transfers
the quantum state from $\ket{0}$ into $\ket{1}$ definitely, while the control parameters evolve
smoothly. Therefore, the pure state protocol is kinematically controllable. But
as shown in FIG. \ref{pures} (b), the main excitation number $N(t)$ is negative, which can
not be feasible  in experiment-setting, so that the reverse engineering protocol is not
dynamically controllable for the pure-state trajectory.

 \begin{figure}[htbp]
\centerline{\includegraphics[width=1.1\columnwidth]{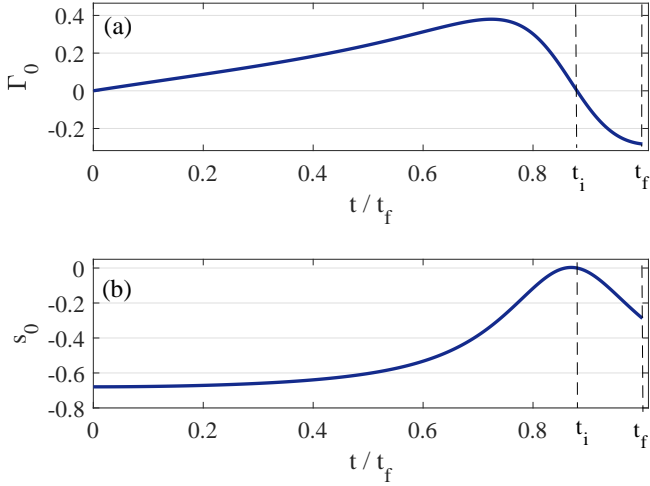}}
\caption{(a) The decay rate $\Gamma_0$ and (b) the Lamb shift  vs the   dimensionless
time $t/t_f$, where $t_f$ is the control pulse length and $t_i$ is the moment where $r_z=0$.
Parameters: $\lambda=0.1\gamma_0,\,\Delta=-0.6792\gamma_0,\,\delta=0.1\gamma_0,\,\Omega_c=
1\gamma_0, t_f=9.1201/\gamma_0$.  We set $\gamma_0=1$ as the unit of $\Gamma_0$ and
$s_0$.}\label{gs}
\end{figure}

Secondly, we show that the population inversion is dynamically controllable if a
mixed-state trajectory of the two-level non-Markovian system is carefully selected. As
we see, the dynamical uncontrollability comes from the negative main excitation number.
We can rewrite the main excitation number as
\begin{eqnarray}
N(t)=-\left(\frac{1}{2}+\frac{{r_{z}}}{{r}^{2}+{r_{z}}^{2}}+\frac{\partial_{t}r^{2}}
{4\Gamma_0\,\left({r}^{2}+{r_{z}}^{2}\right)}\right)\label{men}
\end{eqnarray}
with the length of the Bloch vector $r=\sqrt{r_x^2+r_y^2+r_z^2}$. If the quantum state
is pure, then $\partial_t r^2=0$ and $r=1$, which results in a negative main excitation
number. The population inversion corresponds to the Bloch vector from $r_z(t)=-1$
to $r_z(t_f)=1$. When $r_z$ varies from -1 to 0, the second term in Eq.(\ref{men}) is
negative. Moreover, if $r$ shortens with evolution, the third term in Eq.(\ref{men}) is
also negative. Thus, it can be ensured that the main excitation number is always positive
in  the lower hemisphere of the Bloch sphere. However, in the upper  hemisphere of the
Bloch sphere, i.e., $r_z>0$, the second term in Eq.(\ref{men}) is positive, and $r$ needs
to increase with time, so that the main excitation number can not be always positive in
the evolution. However, the decay rate $\Gamma_0(t)$ is negative at some intermediate
moment. Therefore, we propose following proposal to realize a dynamically controllable population
inversion: (1). We set $t_f$ as the moment where $\Gamma_0$ reaches the negative
maximum for the first time, and label $t_i$ as the moment when $\Gamma_0(t_i)=0$ for
$t_i\in (0, t_f)$, which is illustrated in FIG. \ref{gs} (a). (2). Since $\Gamma_0(t)>0$ for
$t\in (0,t_i)$, we impose that $r_z(t_i)=0$ and $r_z(t)<0$ for $t<t_i$.  In this way,
it is easy to present a positive main excitation number for $t<t_i$ by selecting a mixed trajectory.
(3). Since $\Gamma_0(t)<0$ for $t\in (t_i ,t_f)$, the third term in Eq.(\ref{men}) will be
negative if $ r^2$ keeps increasing. Thus it is possible to present a positive  the main
excitation number if $\vec r(t)\cdot\dot{\vec r} (t)$ increases fast enough.

 \begin{figure}[htbp]
\centerline{\includegraphics[width=1\columnwidth]{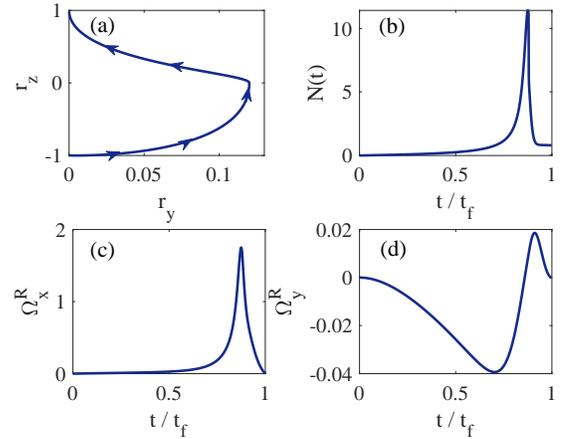}}
\caption{(a) The evolution of quantum state trajectory in the Bloch sphere as a function
of the dimensionless time $t/t_f$. (b) The main excitation number
$N$, (c) the coherent control field $\Omega^{\text{R}}_x$ and (d) the coherent control
field $\Omega^{\text{R}}_y$ as a function of the  dimensionless time $t/t_f$.
Parameters: $\lambda=0.1\gamma_0,\,\Delta=-0.6792\gamma_0,\,\delta=0.1\gamma_0,\,\Omega_c=1
\gamma_0, t_f=9.1201/\gamma_0$.  We set $\gamma_0=1$ as the unit of
$\Omega^{\text{R}}_x$ and $\Omega^{\text{R}}_y$.}\label{ms}
\end{figure}

As an example, we impose the boundary conditions of components of the Bloch vector as follows,
\begin{eqnarray}
\begin{array}{ccc}
  r_y(0)=0, & r_y(t_i)=0.12, & r_y(t_f)=0; \\
  \dot r_y(0)=0, & \dot r_y(t_i)=0, & \dot r_y(t_f)=0; \\
   r_z(0)=-1, & r_z(t_i)=0, & r_z(t_f)=1; \\
\dot r_z(0)=0, & \dot r_z(t_i)=0.4, & \dot r_z(t_f)=1; \\
\end{array}\label{bc}
\end{eqnarray}
and $r_x(t)=0$ for $\forall t$. The reason why we selected the boundary conditions as
Eq.(\ref{bc}) is to eliminate the singular points in the control parameters and obtain the positive
main excitation number. As shown in Eq.(\ref{peq}), $\Omega^{\text{R}}_x$ and
$\Omega^{\text{R}}_y$ have singular points at $t=t_i$  because of $r_z(t_i)=0$. Thus, we
require $\dot r_x(t_i)=\dot r_y(t_i)=0$, and further impose $s_0(t_i)=0$ which is
illustrated in FIG. \ref{gs} (b). On the other hand, if $\vec r\cdot \dot {\vec r}>-
\Gamma_0r_z (r_z+1)$, $N$ will be positive. For $t=t_i$, the positive main excitation
number requires $\dot r_z(t_i)>-\Gamma_0(t_i)$. Thus, the time derivative of $r_z(t_i)$
must be nonzero and finite positive number. Due to $r_z(t_i)=\dot r_y(t_i) =0$, it is not
difficult to verify that $\vec r\cdot \dot {\vec r}=0$, so that the singular point in Eq.(\ref{men})
is also eliminated. At last, the time derivative of $r_z(t_f)$ must be nonzero and
finite positive number, which results in $N(t_f)>0$ (see Eq.(\ref{men})).  To interpolate at
intermediate  times, we assume a polynomial ansatz and consider a piecewise interpolation
with a time break $t_i$. Figure \ref{ms} (a) shows the numerical results of the
quantum state trajectory in the Bloch  sphere, which illustrates that the population is
transferred from $\ket{0}$ into $\ket{1}$ completely. The main excitation number $N$,
the coherent control fields $\Omega^{\text{R}}_x$ and $\Omega^{\text{R}}_y$ are plotted
in FIG.\ref{ms} (b), (c) and (d), respectively. As shown in those figures, the control parameters with
the boundary conditions  of the trajectory Eq.(\ref{bc}) are reasonable, which can be
realized in experimental-settings. Therefore, the population inversion for the two-level
non-Markovian system is dynamical  controllable definitely.

\section{ Conclusions and Discussions}\label{conclusion}

In conclusion, based on the idea of reverse engineering, we have proposed a scheme
to transfer the quantum state of non-Markovian systems along a designable
trajectory in the Hilbert space strictly. For the quantum systems are governed by a
time-convolutionless  master equation, we have presented the analytical expressions
of control parameters, which are the solution of algebraic equations with quantum state trajectories.
Even though the open quantum system suffers the memory effects of the non-Markovian
reservoir (the information backflow or/and the Lamb shift), the quantum state can still
transfer into the target state along designed trajectory strictly. Taking the driven
non-Markovian two-level system as an example, we present concrete control protocol for
both the instantaneous steady state tracking and the population inversion. By elaborately
designing the trajectory of the quantum state, it has been shown that the non-Markovian two-level
system is not only kinematically controllable, but also dynamically controllable.
Since the scheme allows us to maintain system coherence and populations in the
presence of  noises, it may naturally find applications in quantum computing
and quantum memories \cite{Kang2020,Shi2020}. Our scheme can also be applied to numerous
quantum control problems, such as the quantum state preparing \cite{Wu2015},
the quantum measurement \cite{Zheng2020},  and the quantum metrology.

It is meaningful to compare our scheme with the reverse engineering scheme of  the Markovian
quantum systems \cite{ran2020,medina2019}. For the Markovian quantum systems,
it shows that the quantum state is not dynamically controllable \cite{Wu2007,Koch2016}.
For instance, the complete population inversion of two-level systems can not be
realized in the experimental-setting. The population of the excited state is only asymptotically
getting closer to 1, which is discussed in the example of the population inversion
\cite{ran2020}. Due to the information which can flow back into the open two-level systems
\cite{bylicka2017,guarnieri2016}, the complete population inversion for the non-Markovian
dynamics can be realized by  carefully designing the trajectory of the quantum state sweeping
in the Hilbert space. In other words, the non-Markovianity will benefit the quantum
control process.

 This work is supported by National Natural Science Foundation of China (NSFC) under Grants No.
12075050 and 11775048.

\appendix

\section{The Derivation of Eq.(\ref{beq})} \label{appendix}

We begin with the time-convolutionless master equation in the superoperator form Eq.(\ref{meql}).
Taking the density operator vector Eq.(\ref{rho}) into Eq.(\ref{meql})\cite{Schirmer2010},
it yields
\begin{eqnarray}
\partial_{t}r_{i}(t)=\sum_{j=1}^{N^{2}-1}\mathcal{L}_{ij}r_{j}+\mathcal L^0_i,\label{dr}
\end{eqnarray}
where the Liouvillian superoperator can be written in the supermatrix form
\begin{eqnarray}
\hat{\mathcal L}=\sum_{ij=1}^{N^{2}-1}\mathcal L_{i,j} |T_i\rangle\rangle\langle\langle T_j|+
\sum_{i=1}^{N^{2}-1}\mathcal L^0_i|T_i\rangle\rangle\langle\langle T_0|,
\end{eqnarray}
with $\mathcal{L}_{ij}(t)=\text{Tr}[T_{i}^{\dagger}(\hat{\mathcal{L}}[T_{j}])]$ and $\mathcal{L}_{i}^0(t)=\text{Tr}[T_{i}^{\dagger}(\hat{\mathcal{L}}[T_{0}])]$. Here,
the relation $\hat {\mathcal L }^\dagger(t)|T_{0}\rangle\rangle=\langle\langle T_{0}|
\hat {\mathcal L} (t)=0$ has been used. We  divide Liouvillian supermatrix into two parts
$\mathcal{L}_{ij}=\mathcal{C}_{ij}+\mathcal{I}_{ij}$ , where $\mathcal{I}_{ij}$
($\mathcal{C}_{ij}$)  denotes the incoherent (coherent) part of the Liouvillian supermatrix
element $\mathcal{L}_{ij}$.

The coherent part comes from the Hamiltonian  part in the master equation
\begin{eqnarray}
\mathcal{C}_{ij}=-i\:\text{Tr}\left[T_{i}^{\dagger}\,[H(t),T_{j}]\right].
\end{eqnarray}
By substituting Eq.(\ref{hami}) into above equation, we have
\begin{eqnarray}
\mathcal{C}_{ij}=-i\:\sum_{k=0}^{N^{2}-1}c_{k}(t)\,\text{Tr}\left[T_{i}\,[T_{k},T_{j}]\right].
\end{eqnarray}
 Considering the commutator and anti-commutator of the SU($N$) generators
\begin{eqnarray}
[T_{k},T_{j}]&=&i\sum_{m=1}^{N^{2-1}}f_{kjm}T_{m},\\
\left\{ T_{k},T_{j}\right\} &=&\frac{\delta_{kj}}{N}\,I+\sum_{m=1}^{N^{2}-1}d_{kjm}\,T_{m},
\end{eqnarray}
we obtain the coherent part in the Liouvillian supermatrix
\begin{eqnarray}
\mathcal{C}_{ij}&=&\sum_{k=1}^{N^{2}-1}\sum_{m=1}^{N^{2-1}}f_{kjm}c_{k}(t)\,
\text{Tr}[T_{i}T_{m}]\nonumber\\
&=&\sum_{k=1}^{N^{2}-1}\frac{f_{kji}}{2}c_{k}(t),
\end{eqnarray}
where $f_{ijk}$ and $d_{ijk}$ are the structure constants and the d-coefficients of the SU($N$)
Lie algebra, respectively.

For the incoherent part, it can be expressed as
\begin{eqnarray}
\mathcal{I}_{ij}=\text{Tr}\left[T_{i}^{\dagger}\sum_{\alpha}\gamma_{\alpha}\left(2L_{\alpha}
T_{j}L_{\alpha}^{\dagger}-
\left\{ L_{\alpha}^{\dagger}L_{\alpha},T_{j}\right\} \right)\right].
\end{eqnarray}
The Lindblad operators can also be expanded by the SU($N$) Hermitian generators
$\{T_{i}\}_{i=1}^{N^{2}-1}$, i.e
\begin{eqnarray}
L_{\alpha}(t)=\sum_{j=1}^{N^{2}-1}l_{j}^{(\alpha)}(t)\,T_{j}
\end{eqnarray}
with complex coefficients $l_{j}^{(\alpha)}(t)$ , and
\begin{eqnarray}
&&L_{\alpha}^{\dagger}(t)=\sum_{j=1}^{N^{2}-1}l_{j}^{(\alpha)*}(t)\,T_{j},\nonumber\\
&&L_{\alpha}^{\dagger}(t)L_{\alpha}(t)=\sum_{i=0}^{N^{2}-1}e_{k}^{(\alpha)}(t)\,T_{k},
\end{eqnarray}
with $e_{n}^{(\alpha)}=\frac{1}{2}\sum_{i,j=0}^{N^{2}-1}l_{i}^{(\alpha)}(t)l_{j}^{(\alpha)*}(t)\,
\left(i\,f_{ijn}+d_{ijn}\right)$ for $n\neq0$ and $e_{0}^{(\alpha)}=
\sum_{i=0}^{N^{2}-1}\frac{\left|l_{i}^{(\alpha)}(t)\right|^{2}}{2N}$.
Thus, it is easy to obtain
\begin{eqnarray}
&&\text{Tr}\left[T_{i}^{\dagger}\left\{ L_{\alpha}^{\dagger}L_{\alpha},T_{j}\right\} \right]
=e_{0}^{(\alpha)}\,\delta_{ij}+\sum_{k=1}^{N^{2}-1}\frac{d_{kji}}{2}\,e_{k}^{(\alpha)},\nonumber\\
&&\text{Tr}\left[T_{i}^{\dagger}L_{\alpha}T_{j}L_{\alpha}^{\dagger}\right]=\frac{l_{i}^{(\alpha)}
l_{j}^{(\alpha)*}}{4N}+h_{ji}^{(\alpha)},
\end{eqnarray}
with $$h_{ji}^{(\alpha)}=\frac{1}{8}\sum_{p=1}^{N^{2}-1}\sum_{m,n=0}^{N^{2}-1}l_{m}^{(\alpha)}
l_{n}^{(\alpha)*}\left(i\,f_{jnp}+d_{jnp}\right)\left(i\,f_{imp}+d_{imp}\right).$$
Rearranging equations, we finally obtain the incoherent part of the Liouvillian supermatrix
\begin{eqnarray}
\mathcal I_{ij}=\sum_{m,n=0}^{N^{2}-1}\left(\sum_{\alpha}\gamma_{\alpha}l_{m}^{(\alpha)}(t)
l_{n}^{(\alpha)*}(t)\right)s_{mn,ji},
\end{eqnarray}
 with
\begin{eqnarray}
s_{mn,ji}&=&\frac{1}{2N}\left(\delta_{im}\,\delta_{jn}-\delta_{mn}\,\delta_{ij}\right)\nonumber\\
&&+\frac{1}{4}\sum_{k=1}^{N^{2}-1}\left(\left(i\,f_{jnk}+d_{jnk}\right)\left(i\,f_{imk}+
d_{imk}\right)\right.\nonumber\\
&&\left.-\left(i\,f_{mnk}+d_{mnk}\right)\,d_{kji}\right),
\end{eqnarray}
where $|l_{m}^{(\alpha)}(t)|^2=\sum_n l_{m}^{(\alpha)}(t)l_{n}^{(\alpha)*}(t)\delta_{mn}$ has
been used.

The last term in Eq.(\ref{dr}) originates from the expansion of $\mathcal L$ with the basis
$|T_i\rangle\rangle\langle\langle T_0| $. For $j=0$, this term can be
written as
\begin{eqnarray}
\mathcal{L}_{k}^{0}(t)&=&2\text{Tr}\left[T_{k}^{\dagger}\sum_{\alpha}\gamma_{\alpha}(t)
\left(L_{\alpha}(t)L_{\alpha}^{\dagger}(t)-
L_{\alpha}^{\dagger}(t)L_{\alpha}(t)\right)\right]\nonumber\\
&=&\sum_\alpha \gamma_\alpha(t)\left(\sum_{i,j=1}^{N^{2}-1}l_{i}^{(\alpha)}(t)
l_{j}^{(\alpha)*}(t)\right.\nonumber\\
&&\left.\times\left(\left(i\,f_{jik}+d_{jik}\right)-\left(i\,f_{ijk}+d_{ijk}\right)\right)\right)
\end{eqnarray}

\section{A control protocol without $\Omega^R_y$} \label{appendixb}

We also consider a two-level system used in Sec. \ref{example}, whose dynamics is governed
by the non-Markovian master equation Eq.(\ref{nmeq}). Here, we consider
the renormalized control field is real and there is a detuning $\Delta^{\text{R}}$ to the two-level
system. Thus, the Hamiltonian in  Eq.(\ref{nmeq}) can be written as
 \begin{eqnarray}
H_s^{\text{R}}(t)=s_0(t)\sigma_+\sigma_-+\Delta^\text{R}(t)\sigma_z+\Omega^{\text{R}}_x(t)\sigma_x.
 \end{eqnarray}
We still assume that the spectrum density is untunable in experimental-settings. At this time, the
coherent control parameters are $\Omega^\text{R}_x  (t)$ and $\Delta^\text{R}(t)$, while the main excited number
$N(t)$ acts as the incoherent control parameter.  Taking the control parameters and the components
of the Bloch vector into Eq.(\ref{leq}), it yields
  \begin{eqnarray}
\mathrm{\dot{r}_{x}}&=&-(s_{0}+\Delta^\text{R})\,\mathrm{r_{y}}-\left(2\,N+1\right)\,
\Gamma_{0}\,\mathrm{r_{x}},\nonumber\\
\mathrm{\dot{r}_{y}}&=&(s_{0}+\Delta^\text{R})\,\mathrm{r_{x}}-\left(2\,N+1\right)\,
\Gamma_{0}\,\mathrm{r_{y}},\nonumber\\
\mathrm{\dot{r}_{z}}&=&2\,\mathrm{\Omega_{x}^{\text{R}}}\,\mathrm{r_{y}}-2\,
\Gamma_{0}\,\left(\left(2\,N+
1\right)\,\mathrm{r_{z}}+1\right)\!\!.\label{sr1}
 \end{eqnarray}
We can reversely solve Eq.(\ref{sr1}), and obtain
 \begin{eqnarray}
\Omega_{x}^{\text{R}}	&=&	\frac{\left(2\,\Gamma_0+\partial_{t}r_{z}\right)\,
\left(r_{x}^{2}+r_{y}^{2}\right)-\partial_{t}\left(r_{x}^{2}+r_{y}^{2}\right)\,r_{z}}
{2\,\mathrm{r_{y}}\,\left(\mathrm{r}^{2}+\mathrm{r_{z}}^{2}\right)},\nonumber\\
\Delta^{\text{R}}&	=&-s_0+	\frac{r_{x}\,\left(r_{y}\dot{r}_{y}+r_{z}\dot{r}_{z}\right)+
2\,\Gamma_{0}\,r_{x}r_{z}-\dot{r}_{x}\,\left(r_{y}^{2}+2\,r_{z}^{2}\right)}
{\mathrm{r_{y}}\,\left(\mathrm{r_{x}}^{2}+\mathrm{r_{y}}^{2}+2\,\mathrm{r_{z}}^{2}
\right)},\nonumber\\
N	&=&	-\frac{2\,\Gamma_{0}\,\mathrm{r_{z}}+\vec{\mathrm{r}}\cdot\dot{\vec{\mathrm{r}}}
+\mathrm{\Gamma_{0}}\,\left(\mathrm{r}^{2}+\mathrm{r_{z}}^{2}\right)}
{2\,\mathrm{\Gamma_{0}}\,\left(\mathrm{r}^{2}+\mathrm{r_{z}}^{2}\right)}.
 \end{eqnarray}
Thus, we obtain a control protocol without $\Omega^\text{R}_y$. This protocol has advantages
in the population reversion task, because the singular points of control
parameters only appear at $r_y=0$. We may design the trajectory of quantum state away
from points with $r_y=0$. For the completely population reversion, the
initial and final state require $r_y=0$. However, we can set proper boundary
conditions for $r_i$ and $\dot{r}_i$ to eliminate those singular points.




%
%

\end{document}